\documentclass[pdflatex,sn-nature]{sn-jnl}
\usepackage{amsmath,amssymb,graphicx}
\usepackage{etoolbox}
\usepackage{pdfpages}

\apptocmd{\normalsize}{\fontsize{12}{15}\selectfont}{}{}
\apptocmd{\abstractfont}{\fontsize{12}{15}\selectfont\unboldmath}{}{}
\renewcommand{\figurecaptionfont}{\normalfont\normalsize}
\renewcommand{\tablecaptionfont}{\normalfont\normalsize}
\setcitestyle{numbers,square}
\begin{document}
\title{Adiabatic magnetization and the magnetocaloric effect in kagome antiferromagnets}
\author*{\fnm{Katsuhiro} \sur{Morita}} \email{katsuhiro.morita.c7@mso.tohoku.ac.jp}
\affil{Department of Physics, Tohoku University, Sendai, Miyagi 980-8578, Japan}

\abstract{Recent advances in materials synthesis and high-field magnetization measurements have revealed fractional magnetization plateaus in several kagome antiferromagnets. In Cd-kapellasite, pulsed-field measurements have tracked the magnetization up to saturation at approximately 160~T. 
In $\mathrm{YCu_3(OH)_{6.5}Br_{2.5}}$, magnetocaloric measurements have revealed an anomaly near the $1/9$ magnetization plateau. 
To determine how well the uniform kagome model describes these high-field responses at finite temperatures, we used the finite-temperature Lanczos method on a 45-site cluster to calculate the magnetization and spin entropy of the spin-$1/2$ nearest-neighbor kagome Heisenberg antiferromagnet. 
For Cd-kapellasite, accounting for the temperature change during the field sweep by holding the sum of the spin and lattice entropies constant reproduces the gradual approach of the measured magnetization to saturation, particularly above 120~T.
Magnetocaloric data for $\mathrm{YCu}_{3}(\mathrm{OH})_{6.5}\mathrm{Br}_{2.5}$ near 6~K suggest a slight upward bend in $T(H)$ around 20~T, where the $1/9$ plateau appears at lower temperatures.
The uniform model predicts continued cooling with no such bend.}
\maketitle

\section{Introduction}\label{sec:introduction}

Kagome antiferromagnets are paradigmatic geometrically frustrated magnets that host a range of quantum magnetic phenomena, including quantum spin liquids~\cite{Balents2010,Mendels2016,Yan2011,He2017}, magnetization plateaus, and localized magnon states~\cite{Schulenburg2002,ZhitomirskyTsunetsugu2004,Derzhko2004,Schnack2020Magnons}. 
In the spin-$1/2$ kagome Heisenberg antiferromagnet, theory predicts magnetization plateaus at $m=1/9$, $1/3$, $5/9$, and $7/9$~\cite{Nishimoto2013,Picot2016}, where $m=M/M_{\mathrm{sat}}$ is the magnetization normalized to its saturation value. The high-field magnetization curve thus provides a direct probe of the interplay between geometric frustration and quantum fluctuations. 
The nature of the $1/9$ plateau remains under debate~\cite{Fang2023,He2024,Morita2024,Cheng2026,Raikos2026}.
Recent observations of magnetization plateaus in kagome antiferromagnets have stimulated experimental efforts to identify the quantum states underlying these plateaus~\cite{Okuma2019,Suetsugu2024PRL,Jeon2024,Kato2024,Zheng2025PNAS}. Advances in pulsed-field techniques have also made it possible to measure the magnetization of some compounds up to saturation~\cite{Okuma2019}, raising the question of how accurately the ideal kagome model can describe the measured curves.
In distorted kagome antiferromagnets, inequivalent nearest-neighbor exchanges can affect
both the zero-field ground state~\cite{Hering2022} and the magnetization plateaus~\cite{Bodaiji2024}.

Comparisons between high-field magnetization measurements and theory commonly assume that the sample temperature remains constant throughout the field sweep. For field pulses lasting only a few microseconds, however, the sweep can be faster than thermal relaxation between the sample and its surroundings. Under these conditions, the field dependence of the spin entropy can give rise to a change in sample temperature during the sweep through the magnetocaloric effect~\cite{Kohama2010,Kihara2013}. A quantitative interpretation of pulsed-field magnetization measurements therefore requires accounting for the thermal conditions during the sweep~\cite{Morita2022}.

The magnetocaloric effect is particularly relevant to kagome antiferromagnets. 
Geometric frustration and quantum fluctuations give rise to a large number of low-energy states~\cite{Lecheminant1997,Sindzingre2000}, as seen in the
low-temperature specific heat~\cite{Elstner1994,Misguich2005,Chen2018}.
The zero-field spin entropy remains appreciable even at temperatures as low as
$k_{\mathrm B}T/J\sim 0.1$~\cite{Schnack2018}, where $J$ is the nearest-neighbor exchange interaction.
The redistribution of these states by a magnetic field can therefore produce a pronounced magnetocaloric response~\cite{Zhitomirsky2003}. In a material, lattice vibrations also carry entropy. If the spin--phonon relaxation time is short compared with the field-sweep time, the spins and phonons can remain in thermal equilibrium with each other throughout the sweep. If thermal relaxation to the surroundings is slow on the same timescale, the sum of the spin and phonon entropies is approximately conserved, and the field sweep can be treated as adiabatic.

Pulsed-field measurements on the spin-$1/2$ kagome antiferromagnet $\mathrm{YCu_3(OH)_{6.5}Br_{2.5}}$ (YCOB)~\cite{Chen2020YCOB,Liu2022,Chen2025Review} have revealed pronounced changes in sample temperature during the field sweep and a field-dependent magnetic entropy~\cite{Suetsugu2024MCE}. In particular, the entropy anomaly observed near the $1/9$ magnetization plateau illustrates how the magnetocaloric effect can serve as a thermodynamic probe of field-induced quantum states.
Specific-heat measurements also show a feature at the $1/9$ plateau~\cite{Zheng2025}.

 In Cd-kapellasite, high-field magnetization measurements were performed from an initial temperature of $T_0=5~\mathrm{K}$, corresponding to $k_{\mathrm B}T_0/J\sim0.1$. At high fields, the measured magnetization curve deviates substantially from the zero-temperature curve obtained using infinite projected entangled-pair states (iPEPS)~\cite{Okuma2019}. These observations motivate a quantitative study of how the magnetocaloric effect influences the high-field response of kagome antiferromagnets at finite temperatures.

Here we use the finite-temperature Lanczos method (FTLM)~\cite{Jaklic1994,Jaklic2000} to calculate the magnetization and spin entropy of the spin-$1/2$ kagome Heisenberg antiferromagnet as functions of temperature and magnetic field. We use these results to obtain adiabatic magnetization curves and compare our calculations with high-field magnetization and magnetocaloric measurements. The ideal kagome model exhibits a pronounced magnetocaloric effect at low temperatures and high fields, where the temperature change during adiabatic magnetization substantially alters the shape of the magnetization curve. For Cd-kapellasite, including the phonon entropy limits the adiabatic temperature rise and improves agreement with the measured high-field magnetization curve. For YCOB, the ideal model does not reproduce the magnetocaloric response observed near the $1/9$ plateau, suggesting that material-specific effects contribute to this response.

\section{Results and Discussion}\label{sec:results}
\subsection{Spin entropy and adiabatic temperature changes in the uniform kagome antiferromagnet}
\label{sec:entropy}

\begin{figure}[tbp]
    \centering
    \includegraphics[width=100mm]{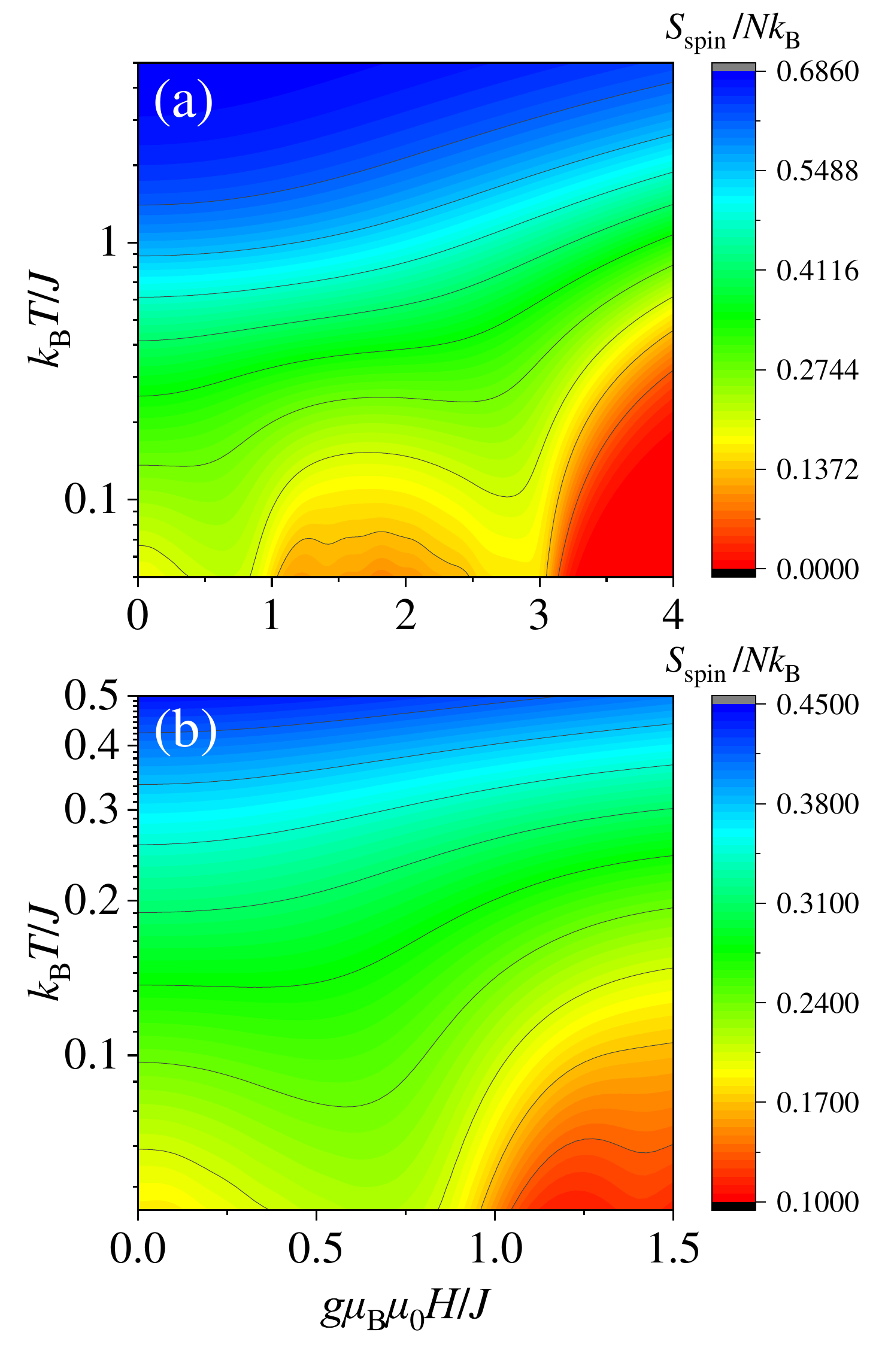}
    \caption{
Spin entropy of the uniform kagome Heisenberg
antiferromagnet.
The color scale shows the spin entropy per site, $S_{\rm spin}/(N k_{\rm B})$,
calculated by the finite-temperature Lanczos method for a 45-site cluster.
Black lines indicate contours of constant spin entropy.
(a) Field and temperature dependence of the spin entropy.
(b) Enlarged view of the low-field, low-temperature region.
The two panels use different color scales.
}
    \label{fig:entropy}
\end{figure}

Figure~\ref{fig:entropy} shows the spin entropy of the spin-$1/2$ nearest-neighbor kagome Heisenberg antiferromagnet as a function of magnetic field and temperature.
The 42-site cluster used in an earlier FTLM study~\cite{Schnack2018}
admits the $m=1/3$ magnetization sector but not those at $m=1/9$, $5/9$, or $7/9$.
To include the magnetization sectors corresponding to all four predicted
plateaus, we use a 45-site cluster for our FTLM calculations.
The horizontal and vertical axes show the reduced field $g\mu_{\mathrm{B}}\mu_0H/J$ and the reduced temperature $k_{\mathrm{B}}T/J$, respectively. The color scale shows the spin entropy per site in units of $k_{\mathrm{B}}$, $S_{\mathrm{spin}}/(N k_{\mathrm{B}})$, and the black lines are isentropes.

At low temperatures, the temperature rises markedly along the isentropes near $g\mu_{\mathrm{B}}\mu_0H/J=1$.
This field lies within the $m=1/3$ plateau, which spans $0.9\lesssim g\mu_{\mathrm{B}}\mu_0H/J\lesssim 1.3$ in previous zero-temperature magnetization calculations~\cite{Nishimoto2013,Hida2001,Cabra2005,Capponi2013}.
Along the lowest-temperature isentrope, the temperature begins to decrease near the upper edge of this plateau.
Along some of the higher-entropy isentropes, the temperature continues to rise up to $g\mu_{\mathrm{B}}\mu_0H/J\simeq 1.7$.

Near the saturation field, the isentropes in Fig.~\ref{fig:entropy}(a) bend sharply toward higher temperatures. Above saturation, increasing the field suppresses spin excitations and reduces the spin entropy at fixed temperature. The temperature must therefore rise to maintain constant spin entropy. Over the temperature range shown, it is difficult to distinguish features in the high-field isentropes that are specific to the $m=5/9$ or $7/9$ plateau.

Figure~\ref{fig:entropy}(b) shows an expanded view of the low-field, low-temperature region. 
The temperature along the isentropes shows no local maximum near $g\mu_{\mathrm{B}}\mu_0H/J=0.4$, where previous zero-temperature calculations place the $m=1/9$ plateau~\cite{Nishimoto2013,Picot2016}.
For example, along the isentrope starting at $k_{\mathrm{B}}T_0/J\simeq 0.10$ in zero field, the temperature continues to decrease gradually through and beyond this field range, reaching a minimum at $g\mu_{\mathrm{B}}\mu_0H/J\simeq 0.6$.

The lower temperature limit in Fig.~\ref{fig:entropy} is $k_{\mathrm{B}}T/J=0.05$. Below this temperature, finite-size effects make it difficult to infer the behavior in the thermodynamic limit from the 45-site results. A detailed comparison of the 36- and 45-site results is given in the Supplementary Information.

In pulsed-field measurements with pulse durations of a few microseconds, the field-sweep time can be much shorter than the thermal relaxation time between the sample and its surroundings. Under these conditions, the magnetocaloric effect can change the sample temperature during the sweep and thereby affect the measured magnetization curve. Figure~\ref{fig:entropy} shows that the temperature changes relatively little at low fields, whereas the isentropes bend appreciably near the $m=1/3$ plateau. 
Several low-temperature isentropes pass through a minimum near the saturation field and then rise sharply with increasing field.

In the next section, we calculate the adiabatic magnetization curve of Cd-kapellasite and compare it with experiment. We assume that the spins and the lattice remain in thermal equilibrium with each other during the field sweep and that the sum of the spin and lattice entropies remains constant.

\subsection{Adiabatic magnetization of Cd-kapellasite}\label{sec:cd}

\begin{figure}[tbp]
    \centering
    \includegraphics[width=100mm]{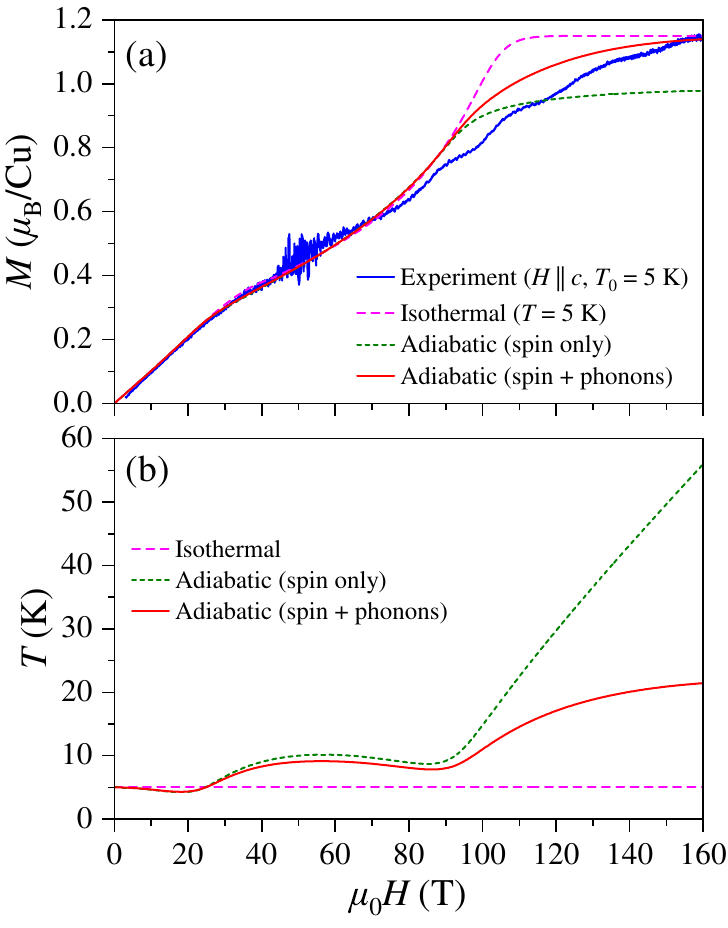}
    \caption{
Magnetization and temperature of Cd-kapellasite
in a pulsed magnetic field.
(a) Magnetization measured on the descending branch of the field
pulse for $H \parallel c$ at an initial temperature of $T_0 = 5$~K (blue),
compared with three calculated curves.
The isothermal calculation holds $T = 5$~K fixed (magenta dashed line).
The adiabatic calculations conserve either the spin entropy alone
(green dashed line) or the sum of the spin and lattice entropies
(red solid line). Both start at $T_0 = 5$~K in zero field.
(b) Calculated temperature as a function of field for the three
conditions in \textbf{(a)}.
Experimental magnetization data are from Ref.~\cite{Okuma2019}.
}
    \label{fig:adiabatic}
\end{figure}

To apply the uniform kagome model to Cd-kapellasite ($\mathrm{CdCu_3(OH)_6(NO_3)_2\cdot H_2O}$), we compared the magnetic susceptibility calculated by FTLM on a 45-site cluster with the susceptibility measured in a field of $1~\mathrm{T}$ applied along the $c$ axis. 
With $g$ fixed at the value of $2.37$ reported in an earlier susceptibility analysis~\cite{Okuma2019}, we obtained $J/k_{\mathrm{B}}=50\,\mathrm{K}$.
The calculated susceptibility reproduces the measured temperature dependence over a broad temperature range. To determine $J$, we focused on temperatures at which finite-size effects in the calculation and low-temperature magnetic ordering in the material~\cite{Okuma2017} have little influence on the susceptibility. Details of the susceptibility analysis and its finite-size dependence are given in the Supplementary Information.

We evaluated the lattice entropy $S_{\mathrm{ph}}(T)$ directly from the analytical expression for the Debye--Einstein model, using parameters from a published fit to the specific heat of Cd-kapellasite~\cite{Akazawa2020}.
The Debye temperature and its weight are $T_{\mathrm{D}}=219.8~\mathrm{K}$ and $A_{\mathrm{D}}=3.052$, and the Einstein temperature and its weight are $T_{\mathrm{E}}=840.3~\mathrm{K}$ and $A_{\mathrm{E}}=4.371$. For the high-field magnetization calculations, we used this lattice entropy and the exchange coupling determined from the susceptibility, $J/k_{\mathrm{B}}=50~\mathrm{K}$.
For the magnetization curves, we used $g=2.30$, corresponding to the measured saturation moment of $1.15\,\mu_{\mathrm{B}}$ per Cu ion for a field along the $c$ axis~\cite{Okuma2019}.

Figure~\ref{fig:adiabatic}(a) compares the calculated magnetization curves with the ultrahigh-field magnetization curve recorded on the descending branch of the field pulse~\cite{Okuma2019}.
 We consider three thermal conditions: an isothermal process at $T=5~\mathrm{K}$, an adiabatic process with constant spin entropy, and an adiabatic process with constant total spin and phonon entropy. Both adiabatic calculations start at $T_0=5~\mathrm{K}$ in zero field. When the lattice contribution is included, we determine the temperature $T(H)$ at each field from

\begin{equation}
S_{\mathrm{spin}}\bigl(T(H),H\bigr)
+ S_{\mathrm{ph}}\bigl(T(H)\bigr)
=
S_{\mathrm{spin}}(T_0,0)
+ S_{\mathrm{ph}}(T_0).
\end{equation}
Here, both entropies refer to the same number $N$ of Cu sites.
The isothermal magnetization curve at $T=5~\mathrm{K}$ agrees reasonably well with experiment up to about $90~\mathrm{T}$. At higher fields, the calculated magnetization increases more rapidly than the measured magnetization. This systematic deviation is consistent with heating during the field pulse, which the isothermal calculation does not capture. When only the spin entropy is held constant, the calculated temperature rises steeply above about $90~\mathrm{T}$, as shown in Fig.~\ref{fig:adiabatic}(b). The resulting thermal excitations strongly suppress the magnetization. Above $140~\mathrm{T}$, agreement with experiment is then poorer than in the isothermal calculation.

When the lattice entropy is included, the phonons take up part of the entropy lost by the spin system as the field increases. This limits the temperature rise and brings the calculated magnetization curve into closer overall agreement with experiment than either the isothermal or the spin-only adiabatic calculation. In particular, the calculation captures the gradual increase in magnetization from $120~\mathrm{T}$ to the saturation field of about $160~\mathrm{T}$. These results show that both the magnetocaloric effect and the lattice contribution to the entropy are important for understanding the ultrahigh-field magnetization of Cd-kapellasite.

Figure~\ref{fig:adiabatic}(b) shows the field dependence of the temperature under the two adiabatic conditions. At $160~\mathrm{T}$, the temperature reaches approximately $55~\mathrm{K}$ when only the spin entropy is conserved, compared with approximately $21~\mathrm{K}$ when phonons are included. This difference accounts for the separation between the two adiabatic magnetization curves at high fields in Fig.~\ref{fig:adiabatic}(a).

Near the $m=1/3$ plateau, the slope of $T(H)$ changes, reflecting the field dependence of the spin entropy. At higher fields, the temperature variation shows no features that can be assigned unambiguously to the $m=5/9$ or $7/9$ plateau. The temperature rise in this region primarily reflects the continuous decrease in spin entropy at fixed temperature as the system approaches saturation.

Although the adiabatic calculation including phonons reproduces the overall shape of the measured magnetization curve, a systematic discrepancy remains between $90$ and $120~\mathrm{T}$. 
The measured differential magnetization exhibits several local minima in this field range. These minima have been interpreted as signatures of additional magnetization plateaus above the $m=1/3$ plateau~\cite{Okuma2019}.
Previous finite-temperature calculations for the $J_1$--$J_2$ kagome model with $|J_2/J_1| \le 0.3$ show no clear magnetization plateaus other than the $m=1/3$ plateau at temperatures relevant here~\cite{Morita2023}. Second-neighbor exchange $J_2$ alone is therefore unlikely to account for the observed high-field anomalies.

A third-neighbor exchange interaction $J_d$ across the hexagon is also unlikely to stabilize plateaus strongly enough to produce clear magnetization anomalies at the temperatures reached during the adiabatic sweep if $J_d$ is sufficiently small compared with $J_1$ and $k_{\mathrm{B}}T \gtrsim |J_d|$.
 One possible origin of the anomalies is a field-induced lattice distortion driven by spin--phonon coupling. Such a distortion has previously been proposed as a mechanism for stabilizing localized magnons~\cite{Richter2004}. In addition to the contribution of lattice entropy included in our calculation, this magnetoelastic effect may contribute to the individual magnetization anomalies.

\subsection{Comparison with the magnetocaloric response of
  \texorpdfstring{$\mathrm{YCu_3(OH)_{6.5}Br_{2.5}}$}{YCu3(OH)6.5Br2.5}}
\label{sec:ycob}

Finally, we compare the isentropes of the uniform nearest-neighbor kagome model with magnetocaloric measurements on YCOB. 
At low temperatures, the sample temperature $T(H)$ passes through a local maximum near 20~T during quasi-adiabatic field sweeps. 
This maximum lies within the field range of the $m=1/9$ magnetization plateau~\cite{Suetsugu2024PRL,Zheng2025PNAS} and has been attributed to the reduction in magnetic entropy associated with the plateau~\cite{Suetsugu2024MCE}.
At about 6~K, Fig.~4(c) of Ref.~\cite{Suetsugu2024MCE} suggests a slight upward bulge in $T(H)$ in the same field range.

Using the exchange coupling $J/k_{\mathrm{B}}\simeq60~\mathrm{K}$ obtained from our analysis of the magnetic susceptibility of YCOB, we find that $6~\mathrm{K}$ corresponds to $k_{\mathrm{B}}T/J\simeq0.10$. Details of the susceptibility analysis are given in the Supplementary Information. As shown in Fig.~\ref{fig:entropy}(b), the temperature along the calculated isentrope with a zero-field temperature of $k_{\mathrm{B}}T_0/J\simeq0.10$ decreases gradually with increasing field. 
The calculated temperature continues to decrease across the field range of the zero-temperature $m=1/9$ plateau near $g\mu_{\mathrm{B}}\mu_0H/J=0.4$ and reaches a minimum at $g\mu_{\mathrm{B}}\mu_0H/J\simeq 0.6$.

Although the uniform nearest-neighbor kagome model exhibits a $1/9$ plateau at zero temperature, it does not reproduce the field dependence of the entropy near the plateau at the experimental temperature. 
A model with three distinct nearest-neighbor exchange couplings has been proposed for the YCOB family~\cite{Hayashi2026}.
This variation in exchange strength may modify the distribution of low-energy excitations near the plateau and thereby contribute to the observed magnetocaloric response.

\section{Conclusions}\label{sec:conclusions}

Recent advances in high-field magnetization measurements have made it possible to follow the magnetization of Cd-kapellasite up to saturation at approximately $160~\mathrm{T}$~\cite{Okuma2019} and to observe the $m=1/9$ magnetization plateau in YCOB at low temperatures~\cite{Suetsugu2024PRL}. Motivated by these experiments, we investigated how well the uniform spin-$1/2$ nearest-neighbor kagome Heisenberg antiferromagnet describes the high-field response of these materials at finite temperatures. Using the spin entropy obtained from finite-temperature Lanczos calculations on a 45-site cluster, we compared the calculated adiabatic magnetization curve of Cd-kapellasite and the isentropes near the $1/9$ plateau in YCOB with the corresponding measurements.

For Cd-kapellasite, we calculated the adiabatic magnetization curve using the exchange coupling $J$ determined from susceptibility measurements at $1~\mathrm{T}$ with $H\parallel c$ and the lattice entropy obtained from a published analysis of specific-heat data~\cite{Akazawa2020}. Including the lattice entropy suppresses the excessive heating predicted by the spin-only adiabatic calculation and reproduces the overall shape of the magnetization curve up to saturation. A systematic discrepancy remains between $90$ and $120~\mathrm{T}$, but above $120~\mathrm{T}$ the calculated magnetization approaches saturation more gradually, in closer agreement with experiment than either the isothermal or the spin-only adiabatic calculation. These results show that accounting for the lattice contribution to the adiabatic temperature change is important for understanding magnetization measurements in pulsed fields.

In YCOB, magnetocaloric data near 6~K suggest a slight upward bend in $T(H)$ around 20~T~\cite{Suetsugu2024MCE}.
At the corresponding initial temperature, $k_{\mathrm{B}}T_0/J\simeq 0.10$, the temperature along the calculated isentrope continues to decrease through the zero-temperature plateau region near $g\mu_{\mathrm{B}}\mu_0H/J=0.4$ and reaches a minimum at $g\mu_{\mathrm{B}}\mu_0H/J\simeq0.6$.
A kagome model incorporating the three distinct nearest-neighbor exchange couplings proposed for the YCOB family~\cite{Hayashi2026} may better describe the overall field dependence of the measured magnetocaloric response. We leave the calculation of the isentropes for this model to future work.

\section{Methods}\label{sec:methods}
\subsection{Model and finite-temperature Lanczos method}\label{sec:ftlm}

We consider the spin-$1/2$ Heisenberg antiferromagnet on the kagome lattice with a uniform nearest-neighbor exchange coupling. The Hamiltonian in a magnetic field is

\begin{equation}
    \mathcal{H} = \mathcal{H}_0 - hQ,
    \qquad
    \mathcal{H}_0 = J \sum_{\langle i,j\rangle}
    \mathbf{S}_i \cdot \mathbf{S}_j,
    \qquad
    Q = \sum_i S_i^z .
\end{equation}

Here, $J>0$ is the antiferromagnetic exchange coupling, $\mathbf{S}_i$ is the spin operator at site $i$, and the sum in $\mathcal{H}_0$ runs over all nearest-neighbor pairs $\langle i,j\rangle$. The operator $Q$ is the $z$ component of the total spin. 
The Zeeman energy is $h=g\mu_{\mathrm{B}}\mu_0H$, where $\mu_0H$ is expressed in tesla. The reduced field $h/J$ is plotted on the horizontal axis of Fig.~\ref{fig:entropy}.

We calculated the magnetization and spin entropy using FTLM~\cite{Jaklic1994,Jaklic2000,Schnack2020FTLM} on a 45-site cluster with periodic boundary conditions. Since $Q$ commutes with $\mathcal{H}_0$, we decomposed the Hilbert space into sectors labeled by the eigenvalue $q$ of $Q$. For a cluster of $N$ sites, $q$ runs from $-N/2$ to $N/2$ in unit steps. We denote the dimension of sector $q$ by $D_q$.

In each sector, we used $R=10$ normalized random starting vectors $|v_{q,r}\rangle$, with $r=1,\ldots,R$. For each starting vector, we applied the standard Lanczos recursion~\cite{Lanczos1950} to $\mathcal{H}_0$ and obtained its tridiagonal representation in the resulting Krylov subspace. We used target Krylov dimensions $L$ ranging from 200 to 400. Diagonalization of the tridiagonal matrix yielded approximate eigenvalues $\varepsilon_{q,r,j}$ and corresponding approximate eigenstates $|\psi_{q,r,j}\rangle$.

The Zeeman term shifts every energy level in sector $q$ by $-hq$. The partition function is therefore approximated by

\begin{equation}
    Z(T,h) \simeq
    \sum_{q=-N/2}^{N/2}
    \frac{D_q}{R}
    \sum_{r=1}^{R} \sum_{j}
    \left| \langle v_{q,r} | \psi_{q,r,j} \rangle \right|^2
    \exp\!\left[-\beta\left(\varepsilon_{q,r,j}-hq\right)\right],
\end{equation}

where $\beta=1/(k_{\mathrm{B}}T)$. The sum over $j$ runs over all eigenstates of the tridiagonal matrix obtained in each Lanczos run.

We calculated the spin entropy from

\begin{equation}
    S_{\mathrm{spin}}(T,h)
    = k_{\mathrm{B}}
    \left[\ln Z(T,h) + \beta E(T,h)\right],
\end{equation}
where $E(T,h)=\langle\mathcal{H}\rangle$ is the internal energy including the Zeeman term. The magnetization normalized to its saturation value is

\begin{equation}
    \frac{M}{M_{\mathrm{sat}}} = \frac{2\langle Q\rangle}{N}.
\end{equation}

We assessed finite-size effects by comparing the 36- and 45-site results. Based on this comparison, we set the lower temperature limit in Fig.~\ref{fig:entropy} to $k_{\mathrm{B}}T/J=0.05$. The geometries of both clusters and the comparison of their results are provided in the Supplementary Information.

\subsection{Lattice entropy and adiabatic magnetization curves}
\label{sec:adiabatic}

For Cd-kapellasite, we obtained the lattice entropy $S_{\mathrm{ph}}(T)$ from the analytical expression for the Debye--Einstein model.
We used the parameters obtained from a fit to the zero-field specific heat between 45 and 300~K:
$T_{\mathrm D}=219.8$~K, $A_{\mathrm D}=3.052$,
$T_{\mathrm E}=840.3$~K, and $A_{\mathrm E}=4.371$~\cite{Akazawa2020}.
The specific heat in that fit was reported per mole of Cu.
The factor $N k_{\mathrm B}$ in Eq.~(6) expresses the lattice entropy on the same $N$-site basis as the spin entropy in Eq.~(4).
With $x_{\mathrm D}=T_{\mathrm D}/T$ and $x_{\mathrm E}=T_{\mathrm E}/T$, the lattice entropy is

\begin{equation}
\begin{aligned}
S_{\mathrm{ph}}(T)
= N k_{\mathrm B}\Bigg\{&
A_{\mathrm D}
\left[4D_3(x_{\mathrm D})
-3\ln\!\left(1-e^{-x_{\mathrm D}}\right)\right] \\
&+3A_{\mathrm E}
\left[\frac{x_{\mathrm E}}{e^{x_{\mathrm E}}-1}
-\ln\!\left(1-e^{-x_{\mathrm E}}\right)\right]
\Bigg\}.
\end{aligned}
\end{equation}
where $D_3(x)$ is the Debye function of order three,

\begin{equation}
    D_3(x)=\frac{3}{x^3}\int_0^x\frac{t^3}{e^t-1}\,\mathrm{d}t .
\end{equation}
The isentropes in Fig.~\ref{fig:entropy} are contours of constant spin entropy. For the adiabatic magnetization of Cd-kapellasite, we also considered the case in which the spins and phonons remain in thermal equilibrium during the field sweep. Starting from a temperature $T_0$ in zero field, we determined the common spin and lattice temperature $T(H)$ at each field by solving

\begin{equation}
    S_{\mathrm{spin}}\bigl(T(H),H\bigr)
    + S_{\mathrm{ph}}\bigl(T(H)\bigr)
    =
    S_{\mathrm{spin}}(T_0,0)
    + S_{\mathrm{ph}}(T_0).
\end{equation}
 We then evaluated the magnetization at $T(H)$ to obtain the adiabatic magnetization curve.

For Fig.~\ref{fig:adiabatic}, we compared the isothermal magnetization at $T=5~\mathrm{K}$ with two adiabatic curves starting at $T_0=5~\mathrm{K}$ in zero field. One adiabatic calculation conserved the spin entropy alone, and the other conserved the sum of the spin and lattice entropies. All three calculations used the exchange coupling $J/k_{\mathrm{B}}=50~\mathrm{K}$ determined from the susceptibility at $\mu_0H=1~\mathrm{T}$ with $H\parallel c$. In the susceptibility analysis, $g$ was fixed at 2.37. For the magnetization curves, we used $g=2.30$, corresponding to the measured saturation moment of $1.15\,\mu_{\mathrm{B}}$ per Cu~\cite{Okuma2019}.

\section*{Acknowledgements}
The author thanks S.~Suetsugu and R.~Okuma for valuable discussions and is particularly grateful to S.~Suetsugu for sharing the magnetic susceptibility data for YCOB.
This work was supported by JSPS KAKENHI Grant Number JP25K07230.
The numerical calculations were performed on the supercomputer Fugaku provided by the RIKEN Center for Computational Science through the HPCI System Research Project (Project ID: hp250573).

\clearpage
\includepdf[pages=-]{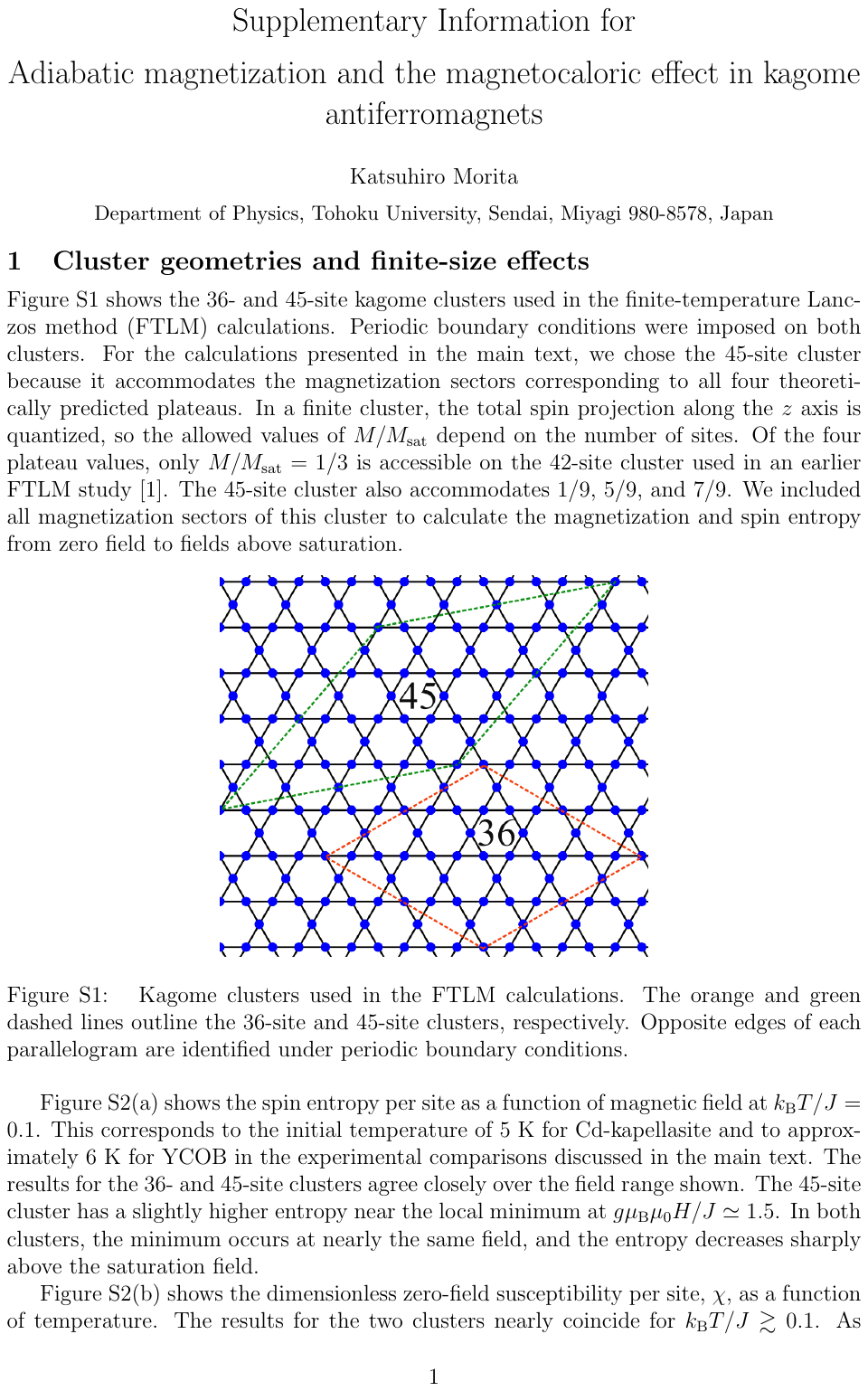}

\end{document}